\documentclass[graybox]{svmult}

\usepackage{mathptmx}       
\usepackage{helvet}         
\usepackage{courier}        
\usepackage{type1cm}        
\usepackage{makeidx}         
\usepackage{graphicx}        
\usepackage{multicol}        
\usepackage[bottom]{footmisc}

\makeindex             

\begin{document}

\title*{UV-IR luminosity functions and stellar mass functions of galaxies
in the Shapley supercluster core}
\titlerunning{UV-IR LFs and SMFs in the SSC} 
\author{ 
A. \,Mecurio \and C. P. \, Haines \and P. \, Merluzzi \and
G. \,Busarello \and R. J. \, Smith \and S. \, Raychaudhury \and
G. P. \, Smith}

\authorrunning{Mercurio, Haines, Merluzzi et al.} 
\institute{
A. Mercurio \and G. Busarello \and P. Merluzzi \at Istituto Nazionale di Astrofisica --
Osservatorio Astronomico di Napoli, Italy,
\\ \email{mercurio, gianni, merluzzi@na.astro.it} 
\and C. P. Haines \and  S. Raychaudhury \and G. P. Smith 
\at  School of Physics and Astronomy,
University of Birmingham, Birmingham B15 2TT
\and R. J. \, Smith
\at Department of
  Physics, University of Durham, Durham DH1 3LE
}

%
%

\maketitle

\vskip-1.0truein

\abstract{We present a panchromatic study of luminosity functions
(LFs) and stellar mass functions (SMFs) of galaxies in the core of the
Shapley supercluster at z=0.048, in order to investigate how the dense
environment affects the galaxy properties, such as star formation (SF)
or stellar masses. We find that while faint-end slopes of optical and
NIR LFs steepen with decreasing density, no environment effect is
found in the slope of the SMFs. This suggests that mechanisms
transforming galaxies in different environments are mainly related to
the quench of SF rather than to mass-loss.  The Near-UV (NUV) and
Far-UV (FUV) LFs obtained have steeper faint-end slopes than the local
field population, while the 24$\mu$m and 70$\mu$m galaxy LFs for the
Shapley supercluster have shapes fully consistent with those obtained
for the local field galaxy population. This apparent lack of
environmental dependence for the infrared (IR) LFs suggests that the
bulk of the star-forming galaxies that make up the observed cluster
IR LF have been recently accreted from the field and have yet to
have their SF activity significantly affected by the cluster
environment.}

\section{Introduction}
\label{intro}

Our panchromatic study of LFs in the Shapley supercluster core (SSC),
from UV to IR wavebands, aims to investigate the relative importance
of the processes that may be responsible for the galaxy
transformations examining in particular the effect of the environment
through the comparison of LFs in regions with different local
densities.

The SSC is constituted by three Abell clusters: A\,3558, A\,3562 and
A\,3556, and two poor clusters SC 1327-312 and SC 1329-313.

The target has been chosen since the most dramatic effects of
environment on galaxy evolution should occur in superclusters, where
the infall and encounter velocities of galaxies are greatest ($>$1 000
km s$^{-1}$), groups and clusters are still merging, and significant
numbers of galaxies will be encountering the dense intra-cluster
medium (ICM) of the supercluster environment for the first time.  This
work is carried out in the framework of the joint research programme
ACCESS
\footnote{http://www.oacn.inaf.it/ACCESS} ({\it A Complete Census of
Star-formation and nuclear activity in the Shapley supercluster},
\cite{MMH10}) aimed at determining the importance of cluster assembly
processes in driving the evolution of galaxies as a function of galaxy
mass and environment within the Shapley supercluster.  We assume
$\Omega_M$ =0.3, $\Omega_{\Lambda}$=0.7 and H$_0$ =70 km s$^{-1}$
Mpc$^{-1}$.

\section{UV-IR Luminosity Functions in different environments}
\label{sec:2}

We analysed wide field $B$- and $R$-band images aquired with the Wide
Field Imager on the 2.2 MPG/ESO telescope (Shapley Optical Survey -
SOS \cite{MMH06}) and $K$-band images obtained at the United Kingdom
Infra-Red Telescope (UKIRT) with the Wide Field Infrared Camera
(WFCAM), covering the whole of the SSC. These data are complemented by
NUV and FUV GALEX data, and Spitzer/MIPS 24$\mu$m and 70$\mu$m imaging
covering essentially the same region ($\sim$ 2.0 deg$^2$) as the
optical and NIR data.  Full details of the observations, data
reduction, and the production of the galaxy catalogues are described
in \cite{MMH06}, \cite{MMH10}, and \cite{HBM10} for the optical, NIR
and UV/IR images, respectively.

\begin{figure}
\centerline{{\resizebox{\hsize}{!}{\includegraphics[angle=-90]{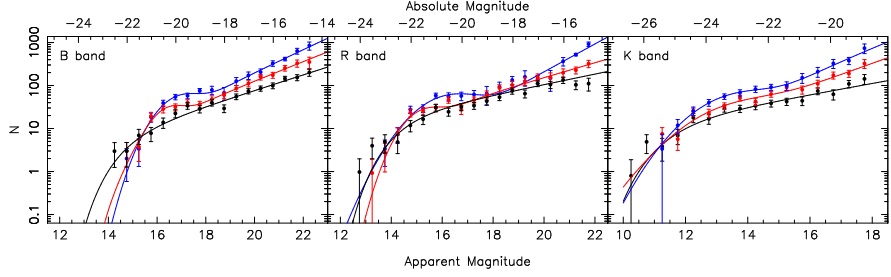}}}}
\caption{Galaxies LFs $B$- (left panel), $R$-band (central panel) and
$K$-band (right panel) in the high- (black), intermediate- (red) and
low-density (blue) environments. Continuous lines represent the G+S
best fit for intermediate- and low-density environments and S best fit
for high-density regions}
\label{LFenv}
\end{figure}

We derive optical and NIR LFs in a 2.0 deg$^2$ area covering the SSC.
In order to investigate the effects of the environment on galaxy
properties, in Fig.~\ref{LFenv} we show and compared the LFs in three
different regions of the supercluster, characterised by high-
($\rho>1.5$; black points, curves), intermediate ($1.0<\rho<1.5$; red),
and low-densities ($0.5<\rho<1.0$ gals arcmin$^{-2}$; blue) of
galaxies. The local density was determined across the $R$-band WFI
mosaic (see \cite{MMH06} for more details). The left and central
panels show,respectively the $B$- and $R$S-band LFs

The optical LFs cannot be described by a single Schechter function (S)
due to the dips apparent at M$^\star$+2 both in $B$ and $R$ bands and
the clear upturn in the counts for galaxies fainter than $B$ and
$R\sim$18\,mag. Instead the sum of a Gaussian and a Schechter function
(G+S), for bright and faint galaxies, respectively, is a suitable
representation of the data.  Furthermore, the slope values becomes
significantly steeper from high- to low-density environments varying
from --1.46$\pm$0.02 to --1.66$\pm$0.03 in $B$ band and from
--1.30$\pm$0.02 to --1.80$\pm$0.04 in $R$ band, being inconsistent at
more than 3$\sigma$ c.l. in both bands. Such a marked luminosity
segregation is related to the behaviour of the red galaxy population:
while red sequence counts are very similar to those obtained for the
global galaxy population, the blue galaxy LFs are well described by
single S and do not vary with the density (see \cite{MMH06}).  This
suggests that mechanisms transforming galaxies in different
environments are mainly related to the quenching of SF.

In Fig.~\ref{LFenv} (right panel), we show the $K$-band LFs of
galaxies in the high- (black), intermediate- (red) and low-density
(blue) regions together with their fits with a Schechter function
(same colour code). Although the fit with a single S function cannot
be rejected in all the three environments, the LFs suggest a bimodal
behaviour due to the presence of an upturn for faint galaxies. To
successfully model these changes in slope and to compare our results
with our optical LFs, we fit our data with a composite G+S. Moreover,
the faint-end slope becomes steeper from high- to low-density
environments varying from --1.33$\pm$0.03 to --1.49$\pm$0.04, being
inconsistent at the 2$\sigma$ confidence level (c.l.)  between high-
and low-density regions.  The observed trend with environment confirm
those observed for the Shapley optical LFs although at lower
significance level dramatic for the Shapley optical LFs.

\begin{figure}
\hbox{
{\resizebox{4.0cm}{!}{\includegraphics{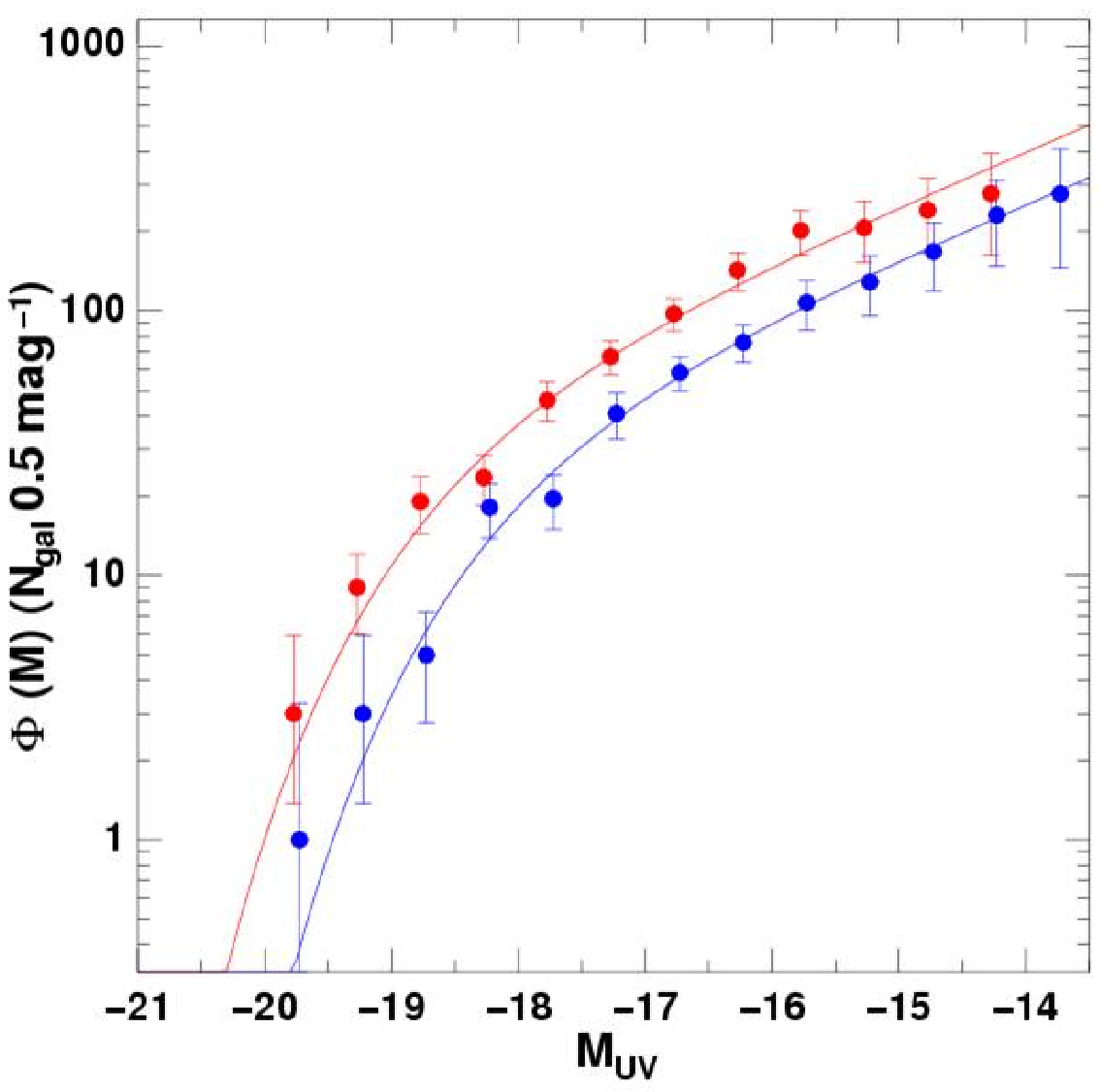}}}
\hfill
{\resizebox{3.8cm}{!}{\includegraphics{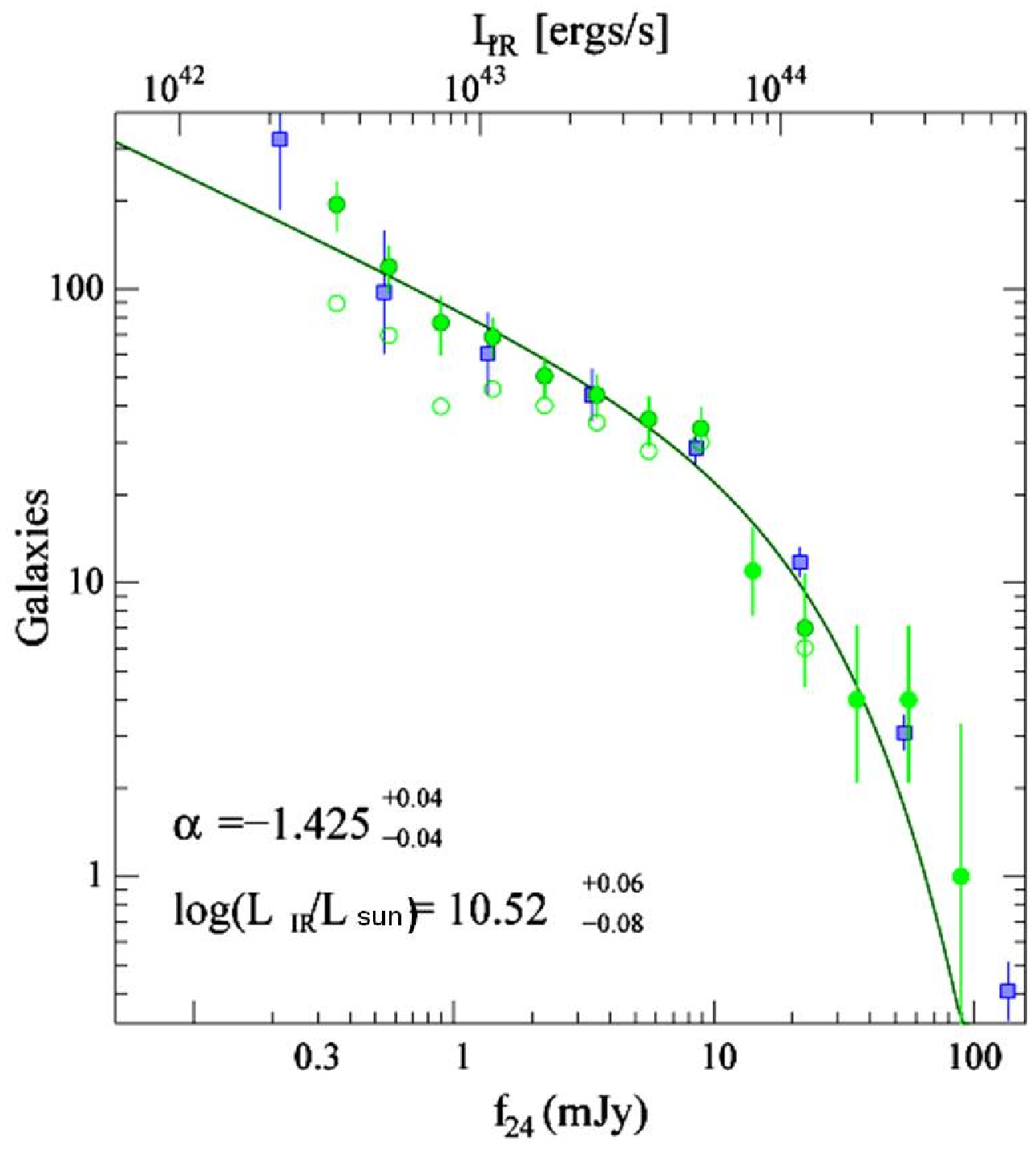}}}
\hfill
{\resizebox{3.8cm}{!}{\includegraphics{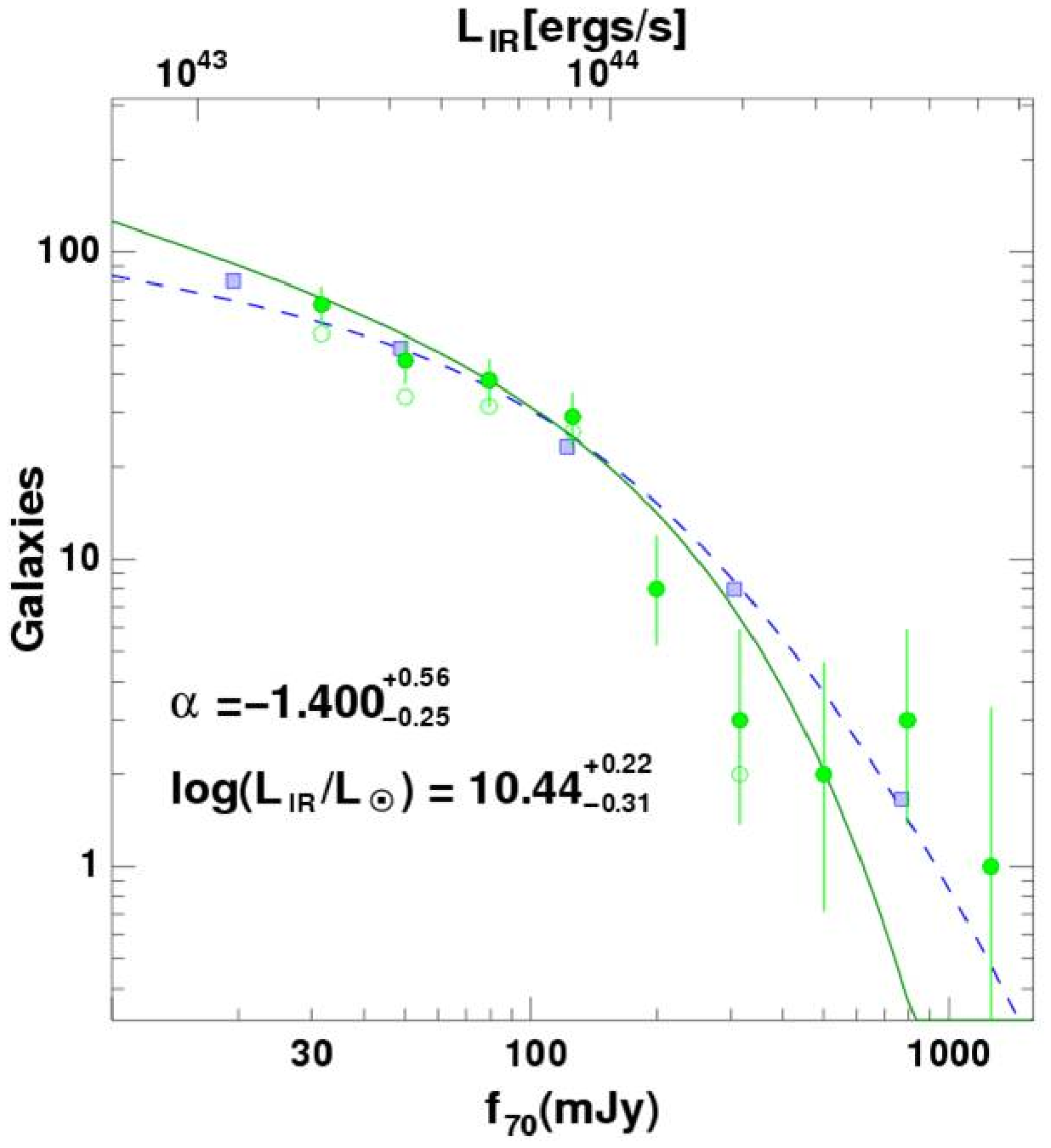}}}
}
\caption{{\it Left panel} FUV (blue) and NUV (red) LFs The solid lines
indicate the best-fitting S functions to the data.{\it Central and
right panel:} 24 $\mu$m and 70 $\mu$m LF (solid green symbols),
respectively. In both panels the contribution due to spectropically
confirmed supercluster members is indicated by open symbols and the
best fitting S is indicated by solid green curve. Blue squares
represent the the field 24$\mu$m LF of \cite{mar07} in the
central panel and in the right panel the field LF of IRAS galaxies from
\cite{wan10}, while the blue dashed curve
indicates the analytic form of the IRAS field IR LF of \cite{tak03}.}
\label{UV_IR_LF}
\end{figure}

From the GALEX data we obtained NUV and FUV LFs (Fig.~\ref{UV_IR_LF},
left panel), which were found to have steeper faint- end slopes
($\alpha=-1.5\pm0.1$) than the local field population
($\alpha=-1.2\pm0.1$) at the $\sim$2 $\sigma$ level, due largely to
the contribution of massive, quiescent galaxies at
M$_{FUV}$$\sim$-16. Using the Spitzer/MIPS 24$\mu$m imaging, we
determined the IR LF of the SSC, finding it well described by a single
Schechter function with log($L^{*}_{IR} / L_{\odot}
)=10.52^{+0.06}_{-0.08}$ and $\alpha_{IR}=-1.49\pm0.04$. We also
presented the first 70$\mu$m LF of a local cluster with Spitzer,
finding it to be consistent with that obtained at 24$\mu$m. The shapes
of the 24$\mu$m and 70$\mu$m LFs were also found to be
indistinguishable from those of the local field population (blue
squares in Fig.~\ref{UV_IR_LF} central and right panel). This apparent
lack of environmental dependence for the shape of the FIR luminosity
function suggests that the bulk of the star-forming galaxies that make
up the observed cluster infrared LF have been recently accreted from
the field and have yet to have their SF activity significantly
affected by the cluster environment. As the SF is quenched via
cluster-related processes, the UV and IR emission drops rapidly,
taking them off the LFs, reducing its normalization but not affecting
its shape.

\section{The galaxy stellar mass functions}
\label{sec:5}

The combined optical and NIR data allow us to derive the distribution
of galaxy stellar masses. The sample we analysed is in the magnitude
range 10$\leq K \leq$18 and refers to the $\sim$2 deg$^2$ area covered
by both the SOS and our $K$-band imaging. The stellar masses of
galaxies belonging to the SSC are estimated by means of stellar
population models by Maraston (\cite{mar05}) with a Salpeter initial
mass function constrained by the observed optical and infrared colours
(see \cite{MMH10} for details).  We use the probability that galaxies
are supercluster members as derived by \cite{HMM06} following
\cite{KB01} in order to estimate and correct for the
foreground/background contamination. We choose that stellar masses of
galaxies belonging to the Shapley supercluster contribute to the
galaxy SMF according to their likelihood of belonging to the SSC.

\begin{figure*}
\centerline{{\resizebox{\hsize}{!}{\includegraphics[angle=-90]{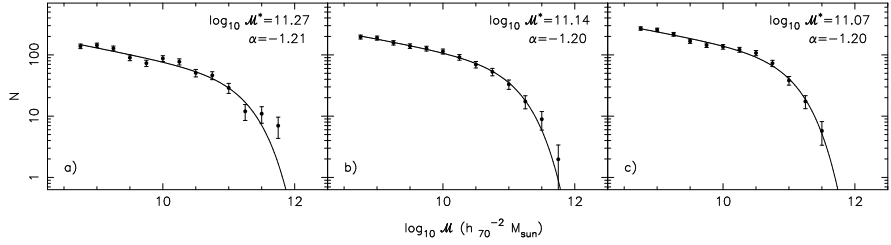}}}}
\caption{The mass function of galaxies in the three cluster regions
corresponding to high- (panel a) , intermediate- (panel b) and
low-density (panel c) environments. In the left, central and right
panel the continuous line represents the fit to the data. In each
panel the best fit value of $\alpha$ and log$_{10}$
$\mathcal{M}^\star$ are reported.}
\label{MFenv}
\end{figure*}

In Fig.~\ref{MFenv} we show the SMF for the different supercluster
environments. Unlike in the case of the optical and NIR LFs no
environmental trend is seen in the slope of the SMFs. On the other
hand, the $\mathcal{M}^\star$ increase from low- to high-density
regions and the excess of high-mass galaxies remains dependent on the
environment.  The different behaviours of LF and of galaxy SMF with
the environment confirm that the mechanisms transforming galaxies in
different environments are mainly related to the quenching of SF rather
than to mass-loss.

\section{Conclusions}
\label{sec:7}

We find that optical and the $K$-band LF faint-end slope becomes
steeper from high- to low-density environments, although the changes
in slope are less dramatic at NIR wavebands indicating that the faint
galaxy population increases in low-density environments. Differently
from the LF no environmental effect is found in the slope of the
SMFs. On the other hand, the $\mathcal{M}^\star$ increase from low- to
high-density regions and the excess of galaxies at the bright-end is
also dependent on the environment.  
These results seem indicate that the physical mechanism responsible
for the transformation of galaxies properties in different environment
are mainly related to the quenching of the SF.  Moreover, the NUV and
FUV LFs obtained have steeper faint-end slopes than the local field
population, while the 24$\mu$m and 70$\mu$m galaxy LFs for the Shapley
supercluster have shapes fully consistent with those obtained for the
Coma cluster and for the local field galaxy population. This apparent
lack of environmental dependence for the shape of IR luminosity
functions suggests that the bulk of the star-forming galaxies that
make up the observed cluster infrared LF have been recently accreted
from the field and have yet to have their SF activity significantly
affected by the cluster environment.

\begin{acknowledgement}

This work was carried out in the framework of the
FP7-PEOPLE-IRSES-2008 project ACCESS. AM acknowledges financial
support from INAF-OAC and the JENAM grant to attend the conference.
\end{acknowledgement}

\end{document}